\documentclass[aip,amsmath,amssymb, reprint]{revtex4-2}
\usepackage{amsmath}
\usepackage{amssymb}
\usepackage{stmaryrd}
\usepackage{graphicx}
\usepackage{esint}
\usepackage{subfigure}
\usepackage{multirow}
\usepackage{mathtools}
\usepackage{xcolor}
\usepackage{gensymb}
\usepackage[normalem]{ulem}
\usepackage{calligra}
\usepackage[T1]{fontenc}

\makeatletter

\newcommand{\beq}{\begin{equation}}
\newcommand{\eeq}{\end{equation}}
\newcommand{\bea}{\begin{eqnarray}}
\newcommand{\eea}{\end{eqnarray}}
\newcommand{\bwt}{\begin{widetext}}
\newcommand{\ewt}{\end{widetext}}
\@ifundefined{textcolor}{}
{%
 \definecolor{BLACK}{gray}{0}
 \definecolor{WHITE}{gray}{1}
 \definecolor{RED}{rgb}{1,0,0}
 \definecolor{GREEN}{rgb}{0,1,0}
 \definecolor{BLUE}{rgb}{0,0,1}
 \definecolor{CYAN}{cmyk}{1,0,0,0}
 \definecolor{MAGENTA}{cmyk}{0,1,0,0}
 \definecolor{YELLOW}{cmyk}{0,0,1,0}
}

\newcommand{\bj}{\mathbf{j}}

\newcommand{\br}{\mathbf{r}}

\newcommand{\bE}{\mathbf{E}}

\newcommand{\bx}{\mathbf{x}}
\newcommand{\by}{\mathbf{y}}

\usepackage{babel}
\makeatother
\usepackage{babel}
\begin{document}

\title{Anisotropic resistivity tensor from disk geometry magneto-conductance}

\author{Oskar Vafek}
\email{vafek@magnet.fsu.edu}
\affiliation{National High Magnetic Field Laboratory, Tallahassee, Florida, 32310, USA}
\affiliation{Department of Physics,
Florida State University, Tallahassee, Florida 32306, USA}

\begin{abstract}
Magneto-transport measurements on two dimensional van der Waals heterostructures have recently shown signatures of uniaxial anisotropy.
Such measurements are almost exclusively performed in a Hall bar geometry which makes it difficult to extract the full resistivity tensor. The goal of this paper is to theoretically analyze anisotropic magneto-conductance in a homogeneous disk geometry and to provide a closed form expression for the electrical potential anywhere on the disk if the current source and drain are located somewhere on the circumference. This expression can then be used to experimentally extract the full conductivity tensor, and by a simple inversion, the full resistivity tensor.
\end{abstract}

\maketitle

\section{Introduction}
Two dimensional van der Waals (vdW) heterostructures host a broad range of interesting physical phenomena\cite{BalentsNatPhys2020}, including anisotropic magnetotransport.
With rare exceptions\cite{KinFaiMak2019,JiaLi}, the transport measurements are performed in a Hall bar geometry, making it difficult to extract the full resistivity tensor particularly if the transport principal axis is misaligned with the current flow. For example, the heterostructures can be subject to an unintentional strain, in which case the misalignment is not directly controlled in an experiment. Moreover, the orientation of the electrical transport principal axis can be carrier concentration (filling) dependent as was recently shown \cite{Xiaoyu} in numerical solutions of the Boltzman equation for twisted bilayer graphene subject to heterostrain, even if the strain tensor and the transport relaxation time are momentum and filling independent. For open Fermi surfaces, the magneto-resistance is expected to grow with the magnetic field $B$ without saturation along one of the principal axis, but to saturate with increasing $B$ along the perpendicular principal axis\cite{Lifshits}. Direct measurement of the full anisotropic resistivity tensor in the vdW heterostructures as a function of filling and $B$ would therefore help in understanding the complex transport phenomena in these materials.

One recent suggestion is to make a ``sunflower'' device \cite{KinFaiMak2019,JiaLi} consisting of a circular disk with thin rectangular petals symmetrically pointing out. In a typical implementation there are $8$ (Ref.\onlinecite{JiaLi}) or $16$ petals\cite{KinFaiMak2019}. In the experiment, the current can be injected along any one of the petals and drained along any other petal. At the same time, the voltage differences can be measured across any remaining pair of petals.
Intuitively, if the source and the drain are $180^{\circ}$ apart, say north and south, and the voltage drop is detected along the side, say northeast and southeast, the resulting resistance will depend on some, possibly complicated, admixture of the components of the conductivity tensor. Holding the relative orientation of the source, the drain and the voltage detection leads fixed, the resistance measurement can be performed for the four petals which are adjacent to the previous set i.e. rotated relative to them by $45^\circ$ for $8$ petals (the rotation would be by $22.5^\circ$ for $16$ petals). If the system has anisotropic conductivity, the measured resistance will be different for the rotated configuration. This rotation can be continued until returning to the original orientation. Such a measurement, as well as a large combination of different source, drain, and probe petal choices, contains information about the anisotropic conductivity tensor. The challenge is to extract this tensor.

In this paper we derive an expression for the voltage at an arbitrary location on the uniform disk of radius $a$ with the current $I$ injected at the source and removed at the drain. The source and drain are placed at an arbitrary pair of points on the boundary. The expression is derived for an arbitrary local conductivity tensor and can be used to extract this tensor and its orientation from the sunflower experiment; resistivity tensor follows from a trivial inversion of conductivity tensor, a $2\times2$ matrix. The extraction can be done as follows: for each source and drain location, there are only $4$ parameters which determine the entire electrical potential profile. They are the two values of the conductivity tensor along the principal axes $\sigma_{\pm}$, the Hall conductivity $\sigma_H$ and the orientation of the principal axes relative to the lab axes; the dependence on $I$ is trivial in the linear $I-V$ regime, it is just an overall scaling factor. The ``sunflower'' measurement yields many more combinations of the pair of voltage probes and source-drain locations, therefore over-constraining the possible values of the $4$ parameters. So, in practice, the $4$ parameters are adjusted to match the measured resistances.

Inside the disk the anisotropic conductivity tensor is assumed to be homogeneous and local, while outside the disk there is no conduction.
Without loss of generality, we choose the coordinate system with the $x$ and $y$ axes aligned with the principal axes and adopt the dyadic product to represent the conductivity tensor. The expression for the electrical potential in the lab coordinate system, rotated in the clockwise sense by an angle $\varphi$ relative to the principal axes coordinate system, can then be easily obtained from a simple axes rotation. The results in both frames are stated later in the introduction.

Thus,
\begin{eqnarray}\label{Eqn:sigma}
\sigma =D(x,y)\left(\sigma_+\hat{\bx}\hat{\bx}+\sigma_-\hat{\by}\hat{\by}+\sigma_H\left(\hat{\bx}\hat{\by}-\hat{\by}\hat{\bx}\right)\right),
\end{eqnarray}
where $D(x,y)=\Theta(a^2-x^2-y^2)$ and $\Theta$ is the Heaviside step function, restricting the conduction to the interior of the circle.
Here $\sigma_{\pm}$ are the two components of the longitudinal conductivity along the principal axes and $\sigma_H$ is the Hall conductivity.
We express the longitudinal conductivities as $\sigma_\pm=\bar{\sigma}\pm\Delta\sigma$ and without loss of generality take the $x$-axis to be along the principal axis with larger resistivity i.e. $\Delta\sigma/\bar{\sigma}<0$.
Note that in the principal axes coordinate system the term $\hat{\bx}\hat{\by}+\hat{\by}\hat{\bx}$ is absent and the Eq.(\ref{Eqn:sigma}) is the most general form of the conductivity tensor in 2D.

The analysis spelled out in the Sec.\ref{sec:analysis} then yields an expression in the form of a rapidly convergent series which can be used to extract the resistivity tensor for a point current source/drain at $\br_{S,D}=a(\cos\theta_{A,B},\sin\theta_{A,B})$ as described above. Equation (\ref{Eqn:V1sd}) in the principal axes coordinate system, and Eq.(\ref{Eqn:V1sdLab}) which transforms it into the lab coordinate system, constitute the main result of the paper. They are stated upfront so that those who do not need all the mathematical details presented in Sec.\ref{sec:analysis} can skip it and continue to Sec.\ref{sec:discussion}.

\begin{figure*}[t]
	\centering
	\subfigure[\label{Fig:isoNoHall}]{\includegraphics[width=0.45\columnwidth]{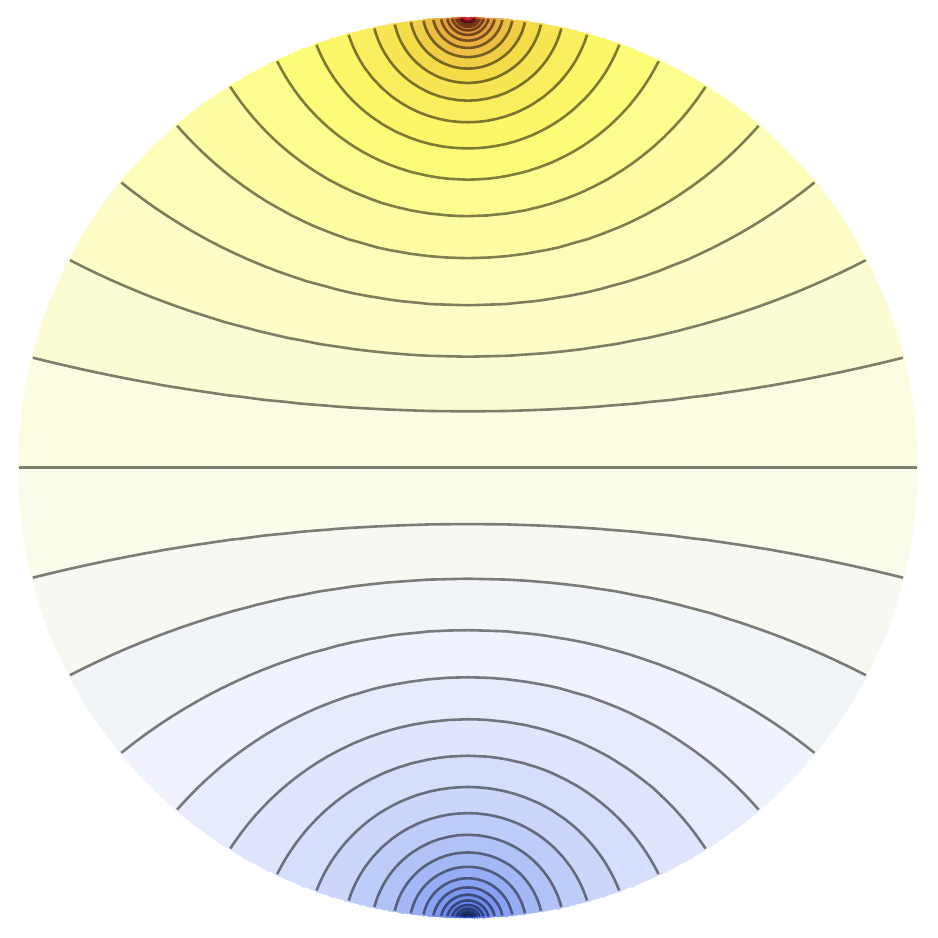}}
\hfill
	\subfigure[\label{Fig:isoHall0pt4}]{\includegraphics[width=0.45\columnwidth]{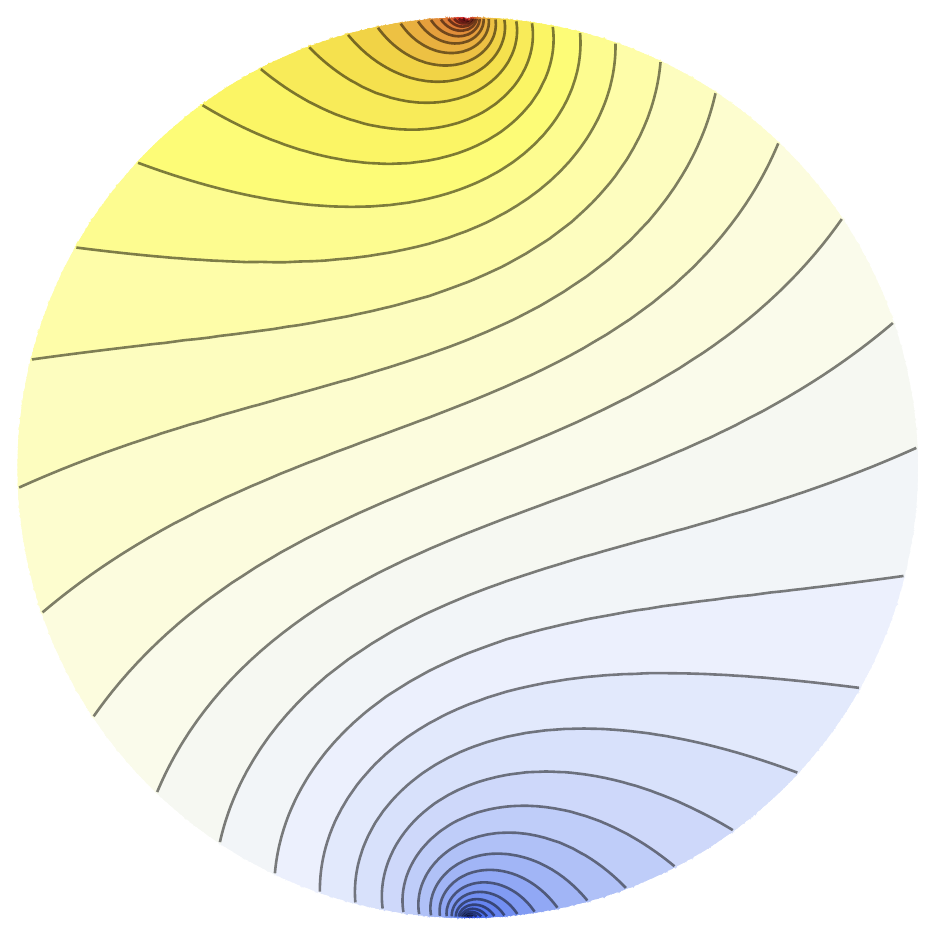}}	
\hfill
	\subfigure[\label{Fig:anisoNoHall}]{\includegraphics[width=0.45\columnwidth]{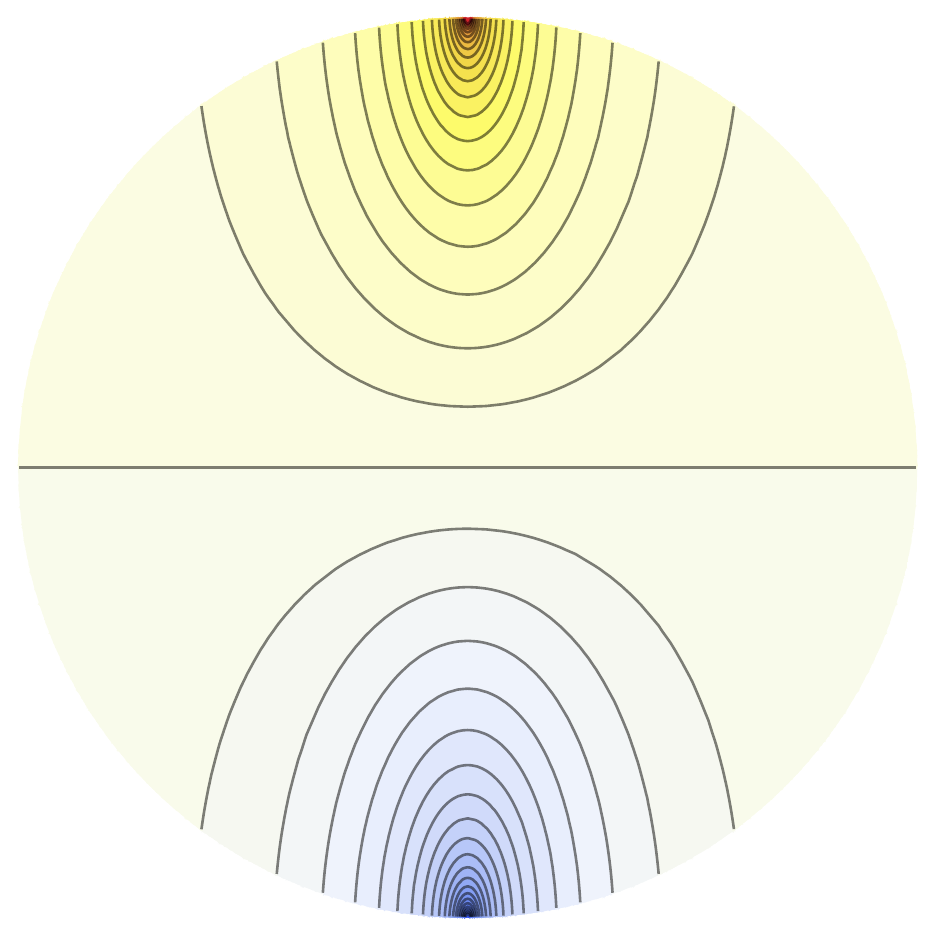}}
\hfill
	\subfigure[\label{Fig:anisoHallOffset}]{\includegraphics[width=0.45\columnwidth]{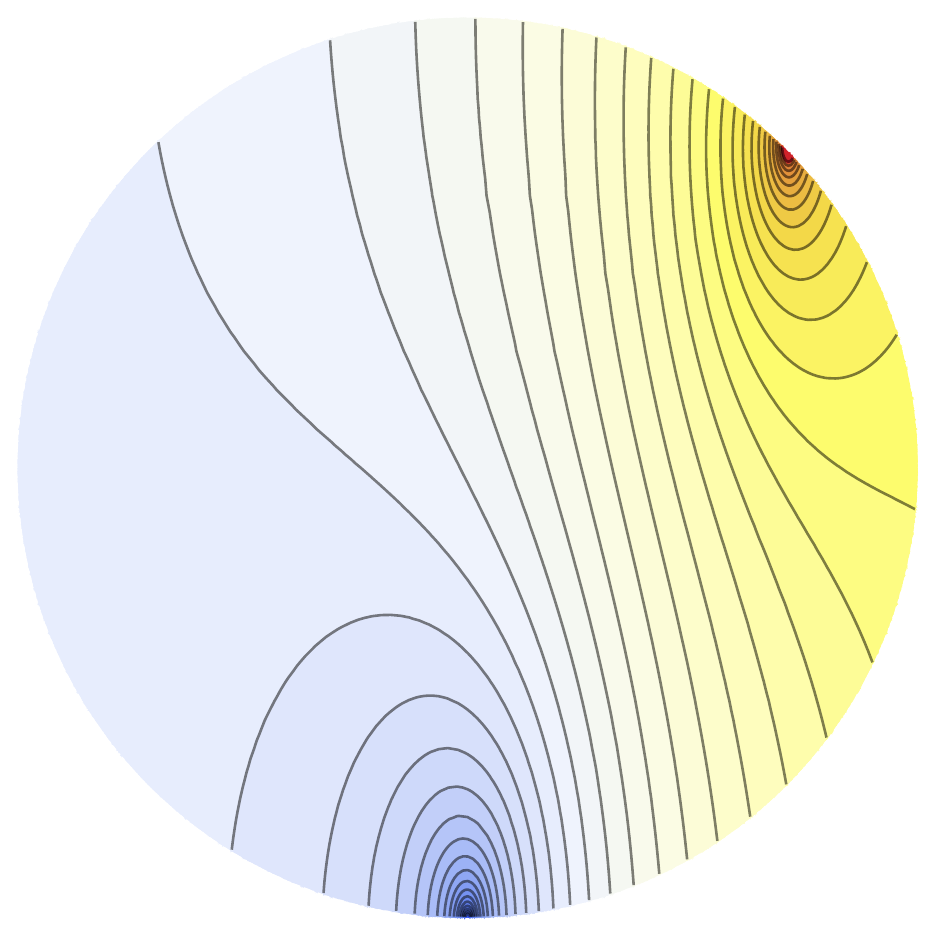}}	
	\caption{Equipotential contours computed using Eq.(\ref{Eqn:V1sd}) in the principal axes coordinate system for isotropic conductivity tensor (a) for $\sigma_H=0$ and (b) for $\sigma_H=0.4\bar\sigma$; in each case the point source is at $\theta_A=\pi/2$ and the point drain at $\theta_B=-\pi/2$. Equipotential contours for anisotropic conductivity tensor for $\sigma_H=0$ and $\Delta\sigma=-0.7\bar\sigma$ for point drain at $\theta_B=-\pi/2$ and (c) point source at $\theta_A=\pi/2$  and (d) $\theta_A=\pi/4$; ten terms in the sum were kept.}
	\label{Fig:potential}
 \end{figure*}

In the principal axes coordinate system, the expression for the electrical potential at $x,y$ reads
\begin{widetext}
\begin{eqnarray}\label{Eqn:V1sd}
&&V(x,y;\br_S,\br_D)=\nonumber\\
&&\frac{I}{\pi}
\frac{\sqrt{\sigma_+\sigma_-}}{\sigma_+\sigma_-+\sigma^2_H}
\left(\sum_{n=0,2,4,\ldots}^{\infty}
\ln\frac{\left|1+e^{-2i\theta_B}\Omega^{2+4n}-e^{-i\theta_B}\frac{Z}{\alpha_+}\Omega^{2n}
\right|}{\left|1+e^{-2i\theta_A}\Omega^{2+4n}-e^{-i\theta_A}\frac{Z}{\alpha_+}\Omega^{2n}
\right|}+
\sum_{n=1,3,5,\ldots}^{\infty}
\ln\frac{\left|1+e^{2i\theta_B}\Omega^{2+4n}-e^{i\theta_B}\frac{Z}{\alpha_+}\Omega^{2n}\right|}{\left|1+e^{2i\theta_A}\Omega^{2+4n}-e^{i\theta_A}\frac{Z}{\alpha_+}\Omega^{2n}\right|}
\right)
\nonumber\\
&+&\frac{I}{\pi}
\frac{\sigma_H}{\sigma_+\sigma_-+\sigma^2_H}\left(
\sum_{n=0,2,4,\ldots}^{\infty}
\arg\left(1+e^{-2i\theta_B}\Omega^{2+4n}-e^{-i\theta_B}\frac{Z}{\alpha_+}\Omega^{2n}\right)-
\arg\left(1+e^{-2i\theta_A}\Omega^{2+4n}-e^{-i\theta_A}\frac{Z}{\alpha_+}\Omega^{2n}\right)
\right.\nonumber\\
&+&\left.\sum_{n=1,3,5,\ldots}^{\infty}
\arg\left(1+e^{2i\theta_B}\Omega^{2+4n}-e^{i\theta_B}\frac{Z}{\alpha_+}\Omega^{2n}\right)-
\arg\left(1+e^{2i\theta_A}\Omega^{2+4n}-e^{i\theta_A}\frac{Z}{\alpha_+}\Omega^{2n}\right)\right).
\end{eqnarray}
\end{widetext}
where the $x,y$ position enters via the complex variable $Z=X+iY=\frac{x}{\sqrt{1+\frac{\Delta\sigma}{\bar{\sigma}}}}+i\frac{y}{\sqrt{1-\frac{\Delta\sigma}{\bar{\sigma}}}}$, and the parameters
$\alpha_{+}=\frac{a}{2}\left(\frac{1}{\sqrt{1+\frac{\Delta\sigma}{\bar{\sigma}}}}+\frac{1}{\sqrt{1-\frac{\Delta\sigma}{\bar{\sigma}}}}\right)$
and  $\Omega=\sqrt{\frac{\sqrt{1-\frac{\Delta\sigma}{\bar\sigma}}-\sqrt{1+\frac{\Delta\sigma}{\bar\sigma}}}{\sqrt{1-\frac{\Delta\sigma}{\bar\sigma}}+\sqrt{1+\frac{\Delta\sigma}{\bar\sigma}}}}
$.
The function $\arg$ is the argument of a complex number.
Note that because $|\Delta\sigma|<\bar{\sigma}$, the parameter $0\leq \Omega<1$ and therefore the above sum converges (the convergence is rapid unless $\Omega$ is very close to $1$). Illustrative contour plots of $V(x,y;\br_S,\br_D)$ for several parameters are shown in the Fig.\ref{Fig:potential}.

Although the above expression is obtained for a point current source/drain, the linearity of the differential equation whose solution it is allows direct determination of the formula for multiple point, as well as spatially extended, current sources/drains. Such formula is presented in the discussion section.

In the lab axes coordinate system which is rotated clockwise by an angle $\varphi$ relative to the principal axes coordinate system, the expression for the electrical potential at $x'$, $y'$ reads
\begin{eqnarray}\label{Eqn:V1sdLab}
&&V^{\text{lab}}(x',y';x'_S,y'_S,x'_D,y'_D)=V(x,y;x_S,y_S,x_D,y_D).
\end{eqnarray}
where $x=x'\cos\varphi + y'\sin\varphi $, $y=y'\cos\varphi - x'\sin\varphi$,
and similarly for $x_{S,D},y_{S,D}$ in terms of $x'_{S,D},y'_{S,D}$.

The paper is organized as follows: Section \ref{sec:analysis} provides mathematical steps to arrive at the expression (\ref{Eqn:V1sd}). First a simple scaling transformation on the $x,y$ variables is performed, turning the disk domain into an ellipse in the $X,Y$ variables. Second a conformal transformation is then performed on the ellipse turning it into an annulus, allowing the final solution. Section III is devoted to discussion and generalization of our results.

\section{Analysis}\label{sec:analysis}
The starting assumption is that Ohm's law holds, i.e.
\begin{eqnarray}
\label{Eqn:ohm}
\bj&=&\sigma\cdot\bE=-\sigma\cdot\nabla V,
\end{eqnarray}
where $\bj$ is the current density, $\bE$ is the electric field and $V$ is the electrical potential, all of which are assumed to be position dependent.
For an idealized point current source and drain, the continuity equation gives
\begin{eqnarray}
\label{Eqn:continuity}
\nabla\cdot \bj&=&I\left(\delta(\br-\br_A)-\delta(\br-\br_B)\right),
\end{eqnarray}
where $I$ is the current, its source is at $\br_A$, its drain at $\br_B$, and $\delta(\br)$ is the Dirac delta function.
Combining Eqs.(\ref{Eqn:sigma}-\ref{Eqn:continuity}) gives
\begin{widetext}
\begin{eqnarray}\label{Eqn:diffeq4V}
-\frac{\partial}{\partial x}\left(D\sigma_+\frac{\partial V}{\partial x}\right)-\frac{\partial}{\partial y}\left(D\sigma_-\frac{\partial V}{\partial y}\right)-\frac{\partial}{\partial x}\left(D\sigma_H\frac{\partial V}{\partial y}\right)+\frac{\partial}{\partial y}\left(D\sigma_H\frac{\partial V}{\partial x}\right)=I\left(\delta(\br-\br_A)-\delta(\br-\br_B)\right).
\end{eqnarray}
\end{widetext}
The solution to the above inhomogeneous linear partial differential equation gives $V$ as a function of $\br$.

Expressing the longitudinal conductivities as
\begin{eqnarray}
\sigma_\pm&=&\bar{\sigma}\pm\Delta\sigma,
\end{eqnarray}
it will be convenient to rescale the coordinate axes according to
\begin{eqnarray}
X&=&\frac{x}{\sqrt{1+\frac{\Delta\sigma}{\bar{\sigma}}}},\\
Y&=&\frac{y}{\sqrt{1-\frac{\Delta\sigma}{\bar{\sigma}}}},
\end{eqnarray}
so that Eq.(\ref{Eqn:diffeq4V}) becomes
\begin{widetext}
\begin{eqnarray}
&&-\left(\frac{\partial}{\partial X}\left(D\frac{\partial V}{\partial X}\right)+\frac{\partial}{\partial Y}\left(D\frac{\partial V}{\partial Y}\right)\right)-\frac{\sigma_H}{\sqrt{\sigma_+\sigma_-}}\left(\frac{\partial}{\partial X}\left(D\frac{\partial V}{\partial Y}\right)-\frac{\partial}{\partial Y}\left(D\frac{\partial V}{\partial X}\right)\right)=\nonumber\\
&&\frac{I}{\sqrt{\sigma_+\sigma_-}}
\left(\delta(X-X_A)\delta(Y-Y_A)-\delta(X-X_B)\delta(Y-Y_B)\right).
\label{Eqn:diffeq4Vscaled}
\end{eqnarray}
\end{widetext}
The new domain, specified by $D\left(\sqrt{1+\frac{\Delta\sigma}{\bar{\sigma}}}X,\sqrt{1-\frac{\Delta\sigma}{\bar{\sigma}}}Y\right)$, is given by
$\Theta(a^2-\left(1+\frac{\Delta\sigma}{\bar{\sigma}}\right)X^2-\left(1-\frac{\Delta\sigma}{\bar{\sigma}}\right)Y^2)$, i.e. it is an ellipse.
If $\Delta\sigma/\bar{\sigma}>0$, the ellipse is elongated along the $Y$-direction, if $\Delta\sigma/\bar{\sigma}<0$, then the ellipse is elongated along the $X$-direction. Without loss of generality we can choose the $x$-axis to be along the axis with larger resistivity, i.e. it will be assumed from now on that
\begin{equation}
\Delta\sigma/\bar{\sigma}<0.
\end{equation}
The equation (\ref{Eqn:diffeq4Vscaled}) can be expressed using complex coordinates
\begin{eqnarray}
Z=X+iY,
\end{eqnarray}
when, after some simplification, it becomes
\begin{widetext}
\begin{eqnarray}
&&-2\left(\frac{\partial}{\partial Z}\left(D\frac{\partial V}{\partial \bar{Z}}\right)+\frac{\partial}{\partial \bar{Z}}\left(D\frac{\partial V}{\partial Z}\right)\right)-\frac{2i\sigma_H}{\sqrt{\sigma_+\sigma_-}}\left(\frac{\partial}{\partial \bar{Z}}\left(D\frac{\partial V}{\partial Z}\right)-\frac{\partial}{\partial Z}\left(D\frac{\partial V}{\partial \bar{Z}}\right)\right)=\nonumber\\
&&\frac{I}{\sqrt{\sigma_+\sigma_-}}
\left(\delta(X-X_A)\delta(Y-Y_A)-\delta(X-X_B)\delta(Y-Y_B)\right).
\label{Eqn:diffeq4VscaledZ}
\end{eqnarray}
\end{widetext}
To avoid confusion, the right-hand-side is kept in terms of the real and imaginary parts of $Z$.
This form makes it clear that inside the ellipse where $D=1$, the solution can be written in terms of a sum of a function of $Z$ and a function of $\bar{Z}$.
The boundary conditions are determined from the right hand side and the derivatives of the boundary function $D$.

\subsection{Zhukovsky conformal mapping of the ellipse to annulus}
\begin{figure*}[t]
	\centering
	\subfigure[\label{Fig:ellipse}]{\includegraphics[width=0.85\columnwidth]{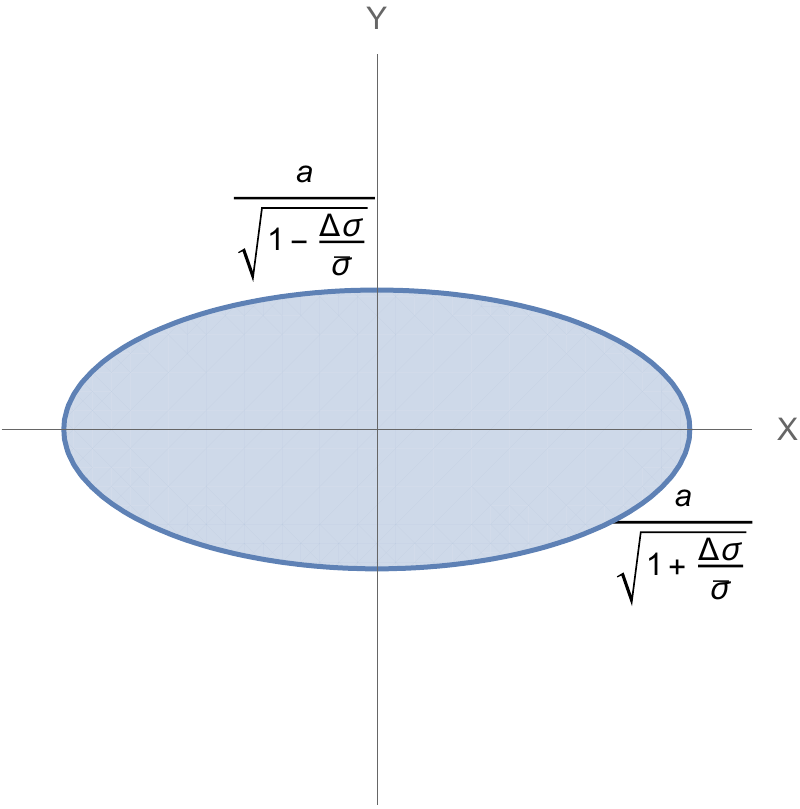}}
\hfill
	\subfigure[\label{Fig:annulus}]{\includegraphics[width=0.85\columnwidth]{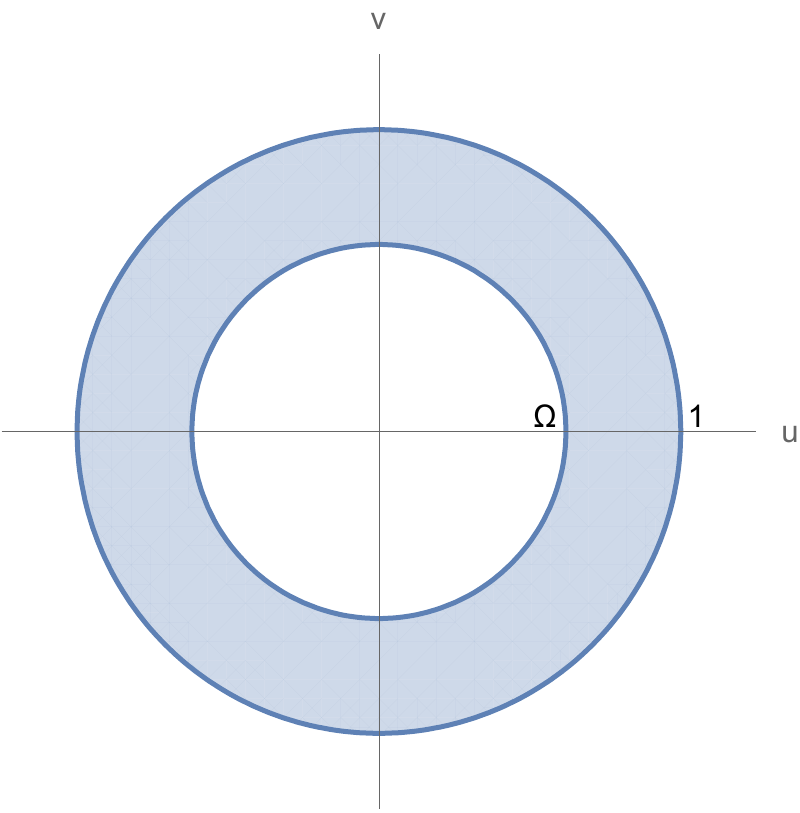}}	
	\caption{For $\Delta\sigma/\bar{\sigma}<0$, (a) the interior of the circular device with radius a maps onto the interior of the ellipse after the coordinate rescaling. (b) The interior of the ellipse in the $X,Y$-plane conformally maps onto the annulus in the $u,v$-plane with unit outer radius and inner radius set by $\Omega$.}
	\label{Fig:Zhukovsky}
 \end{figure*}

It will be convenient to perform a conformal map transforming the boundary of the ellipse to the boundary of the circle.
This can be done using the Zhukovsky transformation
\begin{eqnarray}
\label{Eqn:Zhukovsky}
Z&=&\alpha_+ w+\frac{\alpha_-}{w},\\
w&=&u+iv,
\end{eqnarray}
where $u(X,Y)$ and $v(X,Y)$ are purely real. To determine the coefficients $\alpha_+$ and $\alpha_-$ we demand that
\begin{eqnarray}\label{Eqn:ellipsTocircle}
\left(1+\frac{\Delta\sigma}{\bar{\sigma}}\right)X_0^2+\left(1-\frac{\Delta\sigma}{\bar{\sigma}}\right)Y_0^2&=&a^2,
\end{eqnarray}
implies
\begin{eqnarray}
u_0^2+v_0^2=1,
\end{eqnarray}
i.e. if $X_0$ and $Y_0$ lie on the ellipse, then $u_0$ and $v_0$ are forced to lie on the unit circle.
From Eq.(\ref{Eqn:Zhukovsky}), we have
\begin{eqnarray}\label{Eqn:ellipseANDcircle}
X_0+iY_0&=&\alpha_+(u_0+iv_0)+\alpha_-(u_0-iv_0),
\end{eqnarray}
because, being on unit circle, $1/(u_0+iv_0)=u_0-iv_0$. Therefore,
\begin{eqnarray}
X_0&=&(\alpha_++\alpha_-)u_0,\\
Y_0&=&(\alpha_+-\alpha_-)v_0.
\end{eqnarray}
So, from (\ref{Eqn:ellipsTocircle})
\begin{eqnarray}\label{Eqn:ellipsTocircle2}
&&\left(1+\frac{\Delta\sigma}{\bar{\sigma}}\right)(\alpha_++\alpha_-)^2u^2_0+\left(1-\frac{\Delta\sigma}{\bar{\sigma}}\right)(\alpha_+-\alpha_-)^2v^2_0=a^2,\nonumber\\
\end{eqnarray}
which implies
\begin{eqnarray}\label{Eqn:alphas}
\alpha_{\pm}=\frac{a}{2}\left(\frac{1}{\sqrt{1+\frac{\Delta\sigma}{\bar{\sigma}}}}\pm\frac{1}{\sqrt{1-\frac{\Delta\sigma}{\bar{\sigma}}}}\right).
\end{eqnarray}
This fixes the conformal map.
Having established that the ellipse in the $(X,Y)$-plane maps onto the unit circle in the $(u,v)$-plane, we wish to know where does the interior of the ellipse map. To this end, seek such $w=\Omega$ that would give
\begin{eqnarray}\label{Eqn:alphaPM}
\alpha_+ \Omega&=&\frac{\alpha_-}{\Omega}\in \Re e,
\end{eqnarray}
for $\Delta\sigma/\bar\sigma<0$.
This gives
\begin{eqnarray}
\Omega&=&\sqrt{\frac{\sqrt{1-\frac{\Delta\sigma}{\bar\sigma}}-\sqrt{1+\frac{\Delta\sigma}{\bar\sigma}}}{\sqrt{1-\frac{\Delta\sigma}{\bar\sigma}}+\sqrt{1+\frac{\Delta\sigma}{\bar\sigma}}}}.
\end{eqnarray}
So, letting $w=\Omega e^{i\phi}$ where $\phi$ is the polar angle in the $u,v$-plane and using (\ref{Eqn:alphaPM}) results in
\begin{equation}\label{Eqn:innerCircleMap}
\alpha_+ \Omega e^{i\phi}+\frac{\alpha_-}{\Omega e^{i\phi}}
=
a\frac{\sqrt{-2\frac{\Delta\sigma}{\bar\sigma}}}{\sqrt{1-\left(\frac{\Delta\sigma}{\bar\sigma}\right)^2}}\cos\phi.
\end{equation}
This means that the circle of radius $\Omega$ in $u,v$-plane maps onto the line segment connecting the foci of the ellipse in the $X,Y$-plane.
For $\Delta\sigma/\bar\sigma<0$, the foci lie on the x-axis.
Therefore, the ellipse in $X,Y$-plane, including its interior, maps onto an annulus in the $u,v$-plane with the outer radius $1$ and the inner radius $\Omega$ as illustrated in the Figure \ref{Fig:Zhukovsky}.

Because $Z=f(w)$ i.e. $Z$ is a function of $w$, $w$ is in turn a function of $Z$, i.e. $w=g(Z)$, the left hand side of the differential equation can be written as
\begin{widetext}
\begin{eqnarray}
&&-2\left(\frac{\partial}{\partial Z}\left(D\frac{\partial V}{\partial \bar{Z}}\right)+\frac{\partial}{\partial \bar{Z}}\left(D\frac{\partial V}{\partial Z}\right)\right)-\frac{2i\sigma_H}{\sqrt{\sigma_+\sigma_-}}\left(\frac{\partial}{\partial \bar{Z}}\left(D\frac{\partial V}{\partial Z}\right)-\frac{\partial}{\partial Z}\left(D\frac{\partial V}{\partial \bar{Z}}\right)\right)=\nonumber\\
&&\frac{\partial w}{\partial Z}\frac{\partial \bar{w}}{\partial \bar{Z}}\left(-\left(\frac{\partial}{\partial u}D\frac{\partial V}{\partial u}+\frac{\partial}{\partial v}D\frac{\partial V}{\partial v}\right)+\frac{\sigma_H}{\sqrt{\sigma_+\sigma_-}}\left(
\frac{\partial D}{\partial v}\frac{\partial V}{\partial u}-\frac{\partial D}{\partial u}\frac{\partial V}{\partial v}\right)\right).
\end{eqnarray}
\end{widetext}
Using Cauchy-Riemann conditions, it can be readily shown that
\begin{eqnarray}
\frac{\partial w}{\partial Z}\frac{\partial \bar{w}}{\partial \bar{Z}}=J\left(\frac{u,v}{X,Y}\right),
\end{eqnarray}
where $J\left(\frac{u,v}{X,Y}\right)$ is the Jacobian determinant.
Eq.(\ref{Eqn:diffeq4VscaledZ}) therefore gives
\begin{widetext}
\begin{equation}
-\left(\frac{\partial}{\partial u}D\frac{\partial V}{\partial u}+\frac{\partial}{\partial v}D\frac{\partial V}{\partial v}\right)+\frac{\sigma_H}{\sqrt{\sigma_+\sigma_-}}\left(
\frac{\partial D}{\partial v}\frac{\partial V}{\partial u}-\frac{\partial D}{\partial u}\frac{\partial V}{\partial v}\right)=\frac{I}{\sqrt{\sigma_+\sigma_-}}
\frac{\delta(X-X_A)\delta(Y-Y_A)-\delta(X-X_B)\delta(Y-Y_B)}{J\left(\frac{u,v}{X,Y}\right)}.
\end{equation}
\end{widetext}
But, by the properties of the Dirac $\delta$ function under coordinate transformation, it follows that
\begin{eqnarray}\label{Eqn:diffeq4Vuv}
&&-\left(\frac{\partial}{\partial u}D\frac{\partial V}{\partial u}+\frac{\partial}{\partial v}D\frac{\partial V}{\partial v}\right)+\frac{\sigma_H}{\sqrt{\sigma_+\sigma_-}}\left(
\frac{\partial D}{\partial v}\frac{\partial V}{\partial u}-\frac{\partial D}{\partial u}\frac{\partial V}{\partial v}\right)=\nonumber\\
&&\frac{I}{\sqrt{\sigma_+\sigma_-}}
\left(\delta(u-u_A)\delta(v-v_A)-\delta(u-u_B)\delta(v-v_B)\right),
\end{eqnarray}
where $D=\Theta\left(1-u^2-v^2\right)$.
Now, because $X_{A,B},Y_{A,B}$ lie on the ellipse, $u_{A,B},v_{A,B}$ must lie on the unit circle. From Eq.\ref{Eqn:ellipseANDcircle}
\begin{eqnarray}
\frac{x_{A,B}}{\sqrt{1+\frac{\Delta\sigma}{\bar\sigma}}}&=&\left(\alpha_++\alpha_-\right)u_{A,B},\\
\frac{y_{A,B}}{\sqrt{1-\frac{\Delta\sigma}{\bar\sigma}}}&=&\left(\alpha_+-\alpha_-\right)v_{A,B}.
\end{eqnarray}
Therefore, using Eq.\ref{Eqn:alphas},
\begin{eqnarray}
&&x_{A,B}=a u_{A,B}\\
&&y_{A,B}=a v_{A,B}.
\end{eqnarray}

\subsection{Polar coordinates in the $u,v$-plane}
Switching to polar coordinates in the $u,v$-plane
\begin{eqnarray}
\rho&=&\sqrt{u^2+v^2},\\
\phi&=&\tan^{-1}\frac{v}{u},
\end{eqnarray}
gives
\begin{eqnarray}
\frac{\partial}{\partial u}&=&
\cos\phi\frac{\partial}{\partial \rho}-\frac{\sin\phi}{\rho}\frac{\partial}{\partial \phi},\\
\frac{\partial}{\partial v}&=&
\sin\phi\frac{\partial}{\partial \rho}+\frac{\cos\phi}{\rho}\frac{\partial}{\partial \phi}.
\end{eqnarray}
Therefore, the derivatives of the boundary function are
\begin{eqnarray}
\frac{\partial D}{\partial u}&=&-\cos\phi\delta(\rho-1),\\
\frac{\partial D}{\partial v}&=&-\sin\phi\delta(\rho-1),
\end{eqnarray}
and the differential equation (\ref{Eqn:diffeq4Vuv}) becomes
\begin{widetext}
\begin{eqnarray}\label{Eqn:VuvPolarCase1}
&&-\left(\frac{\partial}{\partial \rho}\left(D\rho\frac{\partial V}{\partial \rho}\right)+\frac{D}{\rho}\frac{\partial^2 V}{\partial \phi^2}\right)+\frac{\sigma_H\delta(\rho-1)}{\sqrt{\sigma_+\sigma_-}}
\frac{\partial V}{\partial \phi}=\frac{I}{\sqrt{\sigma_+\sigma_-}}\delta(\rho-1)
\left(\delta(\phi-\theta_A)-\delta(\phi-\theta_B)\right).
\end{eqnarray}

\subsection{Homogeneous solution and the boundary conditions}
A general solution of Eq.(\ref{Eqn:VuvPolarCase1}) for $\rho<1$ where the terms containing $\delta(\rho-1)$ vanish can be written as
\begin{eqnarray}\label{EqnVcase1}
V(\rho,\phi)&=&\sum_{m=1}^{\infty}\left(A_{m}\left(\frac{\rho^m}{\Omega^m}+\frac{\Omega^m}{\rho^{m}}\right)\cos m\phi+
B_{m}\left(\frac{\rho^m}{\Omega^m}-\frac{\Omega^m}{\rho^{m}}\right)\sin m\phi\right).
\end{eqnarray}
This form satisfies the homogeneous differential equation and is continuous and differentiable across the line cut joining the foci.
To see this, notice that the points on the circle of radius $\Omega$ in the $u,v$-plane map onto the line segment joining the foci $X\in (-2\alpha_+\Omega,2\alpha_+\Omega)$, $Y=0$, as we saw in the Eq.\ref{Eqn:innerCircleMap}. Therefore, the points on the inner circle in the $u,v$ plane which are related by the mirror reflection about the $v=0$ axis should be identified as the same points. In other words, $\rho=\Omega$ and $\phi$, and $\rho=\Omega$ and $-\phi$ map onto the same physical point in the $X,Y$ and therefore $x,y$ plane. We therefore want the potential at $\Omega^+$ and $\phi$ to either be the same at $-\phi$ which is accomplished by $\left(\frac{\rho^m}{\Omega^m}+\frac{\Omega^m}{\rho^{m}}\right)\cos m\phi$, or we want it to vanish at $\Omega^+$ with a continuous slope. Vanishing at $\Omega$ is accomplished by $\frac{\rho^m}{\Omega^m}-\frac{\Omega^m}{\rho^{m}}$, and the reason why only $\sin m\phi$ can multiply it is that multiplying it by $\cos m\phi$ would introduce a cusp across the line segment.

Integrating both sides of Eq.\ref{Eqn:VuvPolarCase1} over an infinitesimal interval straddling $\rho=1$ gives the boundary condition
\begin{eqnarray}\label{Eqn:Vbc}
\frac{\partial V}{\partial \rho}|_{\rho=1}+\frac{\sigma_H}{\sqrt{\sigma_+\sigma_-}}
\frac{\partial V}{\partial \phi}|_{\rho=1}=\frac{I}{\sqrt{\sigma_+\sigma_-}}
\left(\delta(\phi-\theta_A)-\delta(\phi-\theta_B)\right).
\end{eqnarray}
Substituting Eq.\ref{EqnVcase1} into the above results in
\begin{eqnarray}
\frac{\partial V(\rho,\phi)}{\partial\rho}|_{\rho=1}&=&\sum_{m=1}^{\infty}\left(
mA_{m}\left(\frac{1}{\Omega^m}-\Omega^m\right)\cos m\phi+
mB_{m}\left(\frac{1}{\Omega^m}+\Omega^m\right)\sin m\phi\right),\\
\frac{\partial V(\rho,\phi)}{\partial\phi}|_{\rho=1}&=&
\sum_{m=1}^{\infty}\left(-mA_{m}\left(\frac{1}{\Omega^m}+\Omega^m\right)\sin m\phi+
mB_{m}\left(\frac{1}{\Omega^m}-\Omega^m\right)\cos m\phi\right),
\end{eqnarray}
and the differential equation (\ref{Eqn:VuvPolarCase1}) becomes
\begin{eqnarray}
&&\sum_{m=1}^{\infty}\left(
mA_{m}\left(\frac{1}{\Omega^m}-\Omega^m\right)\cos m\phi+
mB_{m}\left(\frac{1}{\Omega^m}+\Omega^m\right)\sin m\phi\right)+\nonumber\\
&+&\frac{\sigma_H}{\sqrt{\sigma_+\sigma_-}}
\sum_{m=1}^{\infty}\left(-mA_{m}\left(\frac{1}{\Omega^m}+\Omega^m\right)\sin m\phi+
mB_{m}\left(\frac{1}{\Omega^m}-\Omega^m\right)\cos m\phi\right)=\nonumber\\
&&\frac{I}{\sqrt{\sigma_+\sigma_-}}
\frac{1}{\pi}\sum_{m=1}^{\infty}\left(\cos m\phi\left(\cos m\theta_{A}-\cos m\theta_{B}\right)+\sin m\phi\left(\sin m\theta_{A}-\sin m\theta_{B}\right)\right),
\end{eqnarray}
where the following identity was used for the right hand side
\begin{eqnarray}
\delta\left(\phi-\theta_{A,B}\right)&=&\frac{1}{2\pi}\sum_{m=-\infty}^{\infty}e^{im\phi}e^{-im\theta_{A,B}}
\\
&=&\frac{1}{2\pi}+\frac{1}{\pi}\sum_{m=1}^{\infty}\left(\cos m\phi\cos m\theta_{A,B}+\sin m\phi\sin m\theta_{A,B}\right).
\end{eqnarray}
Matching the coefficients of $\cos m\phi$ and $\sin m\phi$ and solving for $A_{m}$ and $B_{m}$ gives
\begin{eqnarray}
&&A_{m}=\frac{I}{\sqrt{\sigma_+\sigma_-}}
\frac{1}{\pi}\frac{1}{1+\frac{\sigma^2_H}{\sigma_+\sigma_-}}\frac{1}{m}\left(
\frac{\cos m\theta_{A}-\cos m\theta_{B}}{\Omega^{-m}-\Omega^m}
-\frac{\sigma_H}{\sqrt{\sigma_+\sigma_-}}\frac{\sin m\theta_{A}-\sin m\theta_{B}}{\Omega^{-m}+\Omega^m}\right)\\
&&B_{m}=\frac{I}{\sqrt{\sigma_+\sigma_-}}
\frac{1}{\pi}\frac{1}{1+\frac{\sigma^2_H}{\sigma_+\sigma_-}}\frac{1}{m}\left(
\frac{\sigma_H}{\sqrt{\sigma_+\sigma_-}}\frac{\cos m\theta_{A}-\cos m\theta_{B}}{\Omega^{-m}-\Omega^m}
+\frac{\sin m\theta_{A}-\sin m\theta_{B}}{\Omega^{-m}+\Omega^m}\right).
\end{eqnarray}
Thus,
\begin{eqnarray}\label{Vcase1Series0}
&&V(\rho,\phi)=\frac{I}{\sqrt{\sigma_+\sigma_-}}
\frac{1}{\pi}\frac{1}{1+\frac{\sigma^2_H}{\sigma_+\sigma_-}}\sum_{m=1}^{\infty}
\frac{1}{m}\left(
\frac{\cos m\theta_{A}-\cos m\theta_{B}}{\Omega^{-m}-\Omega^m}
-\frac{\sigma_H}{\sqrt{\sigma_+\sigma_-}}\frac{\sin m\theta_{A}-\sin m\theta_{B}}{\Omega^{-m}+\Omega^m}\right)
\left(\frac{\rho^m}{\Omega^m}+\frac{\Omega^m}{\rho^{m}}\right)\cos m\phi\nonumber\\
&+&\frac{I}{\sqrt{\sigma_+\sigma_-}}
\frac{1}{\pi}\frac{1}{1+\frac{\sigma^2_H}{\sigma_+\sigma_-}}\sum_{m=1}^{\infty}
\frac{1}{m}\left(
\frac{\sin m\theta_{A}-\sin m\theta_{B}}{\Omega^{-m}+\Omega^m}+\frac{\sigma_H}{\sqrt{\sigma_+\sigma_-}}\frac{\cos m\theta_{A}-\cos m\theta_{B}}{\Omega^{-m}-\Omega^m}
\right)
\left(\frac{\rho^m}{\Omega^m}-\frac{\Omega^m}{\rho^{m}}\right)\sin m\phi\nonumber\\
&-&(A\rightarrow B).
\end{eqnarray}

\subsection{Summing over the angular momenta}
The sum over $m$ converges slowly. In order to convert it into a rapidly convergent sum, we first Taylor expand the denominators involving $\Omega^m$ and $\Omega^{-m}$, in powers of $\Omega$ as
\begin{eqnarray}\label{Vcase1Series}
&&V(\rho,\phi)=
\frac{I}{\sqrt{\sigma_+\sigma_-}}
\frac{1}{\pi}\frac{1}{1+\frac{\sigma^2_H}{\sigma_+\sigma_-}}\sum_{n=0}^{\infty}\sum_{m=1}^{\infty}
\frac{\left(\Omega^{1+2n}\right)^m}{m}\left(
\cos m\theta_{A}
-(-1)^n\frac{\sigma_H}{\sqrt{\sigma_+\sigma_-}}\sin m\theta_{A}\right)
\left(\frac{\rho^m}{\Omega^m}+\frac{\Omega^m}{\rho^{m}}\right)\cos m\phi\nonumber\\
&+&\frac{I}{\sqrt{\sigma_+\sigma_-}}
\frac{1}{\pi}\frac{1}{1+\frac{\sigma^2_H}{\sigma_+\sigma_-}}\sum_{n=0}^{\infty}\sum_{m=1}^{\infty}
\frac{\left(\Omega^{1+2n}\right)^m}{m}\left(
(-1)^n\sin m\theta_{A}+\frac{\sigma_H}{\sqrt{\sigma_+\sigma_-}}\cos m\theta_{A}
\right)
\left(\frac{\rho^m}{\Omega^m}-\frac{\Omega^m}{\rho^{m}}\right)\sin m\phi\nonumber\\
&-&(A\rightarrow B).
\end{eqnarray}
Then the resulting sum over $m$ is related to the geometric series by integration, and since $\Omega<1$, the sum over $n$ will be rapidly convergent. Therefore, for $C=A,B$, we have
\begin{eqnarray}
\sum_{m=1}^\infty e^{im\theta_C}e^{im\phi}\frac{\Omega^{m(1+2n)}}{m}\left(\frac{\rho^m}{\Omega^m}\pm \frac{\Omega^m}{\rho^m} \right)
&=&-\ln\left(1-e^{i\theta_C}\frac{w}{\Omega}\Omega^{1+2n}\right)\mp \ln\left(1-e^{i\theta_C}\frac{\Omega}{\bar w}\Omega^{1+2n}\right).
\end{eqnarray}
Adding longitudinal and Hall contributions finally gives Eq.(\ref{Eqn:V1sd}).
\end{widetext}
\section{Discussion}\label{sec:discussion}
Because the differential equation is linear, it is straightforward to generalize the expression derived above to the case with multiple sources and drains.
In such a case the continuity equation reads
\begin{equation}
\nabla\cdot \bj=\sum_{j=1}^{n_S}I^{S}_j\delta(\br-\br_{A,j})-\sum_{j=1}^{n_D}I^{D}_j\delta(\br-\br_{B,j}),
\end{equation}
where $n_{S}$ is the number of point sources and $n_D$ is the number of point drains, and $\sum_{j=1}^{n_S}I^{S}_j=\sum_{j=1}^{n_D}I^{D}_j=I$.
The resulting expression is
\begin{widetext}
\begin{eqnarray}\label{Eqn:Vmanysd}
&&V(x,y;\{\br_{A,j}\},\{\br_{B,j}\})=\nonumber\\
&&\frac{1}{\pi}
\frac{\sqrt{\sigma_+\sigma_-}}{\sigma_+\sigma_-+\sigma^2_H}\sum_{j=1}^{n_D}I^{D}_j
\left(\sum_{n=0,2,4,\ldots}^{\infty}
\ln\left|1+e^{-2i\theta_{B,j}}\Omega^{2+4n}-e^{-i\theta_{B,j}}\frac{Z}{\alpha_+}\Omega^{2n}
\right|+
\sum_{n=1,3,5,\ldots}^{\infty}
\ln\left|1+e^{2i\theta_{B,j}}\Omega^{2+4n}-e^{i\theta_{B,j}}\frac{Z}{\alpha_+}\Omega^{2n}\right|\right)
\nonumber\\
&-&\frac{1}{\pi}
\frac{\sqrt{\sigma_+\sigma_-}}{\sigma_+\sigma_-+\sigma^2_H}\sum_{j=1}^{n_S}I^{S}_j
\left(\sum_{n=0,2,4,\ldots}^{\infty}
\ln\left|1+e^{-2i\theta_{A,j}}\Omega^{2+4n}-e^{-i\theta_{A,j}}\frac{Z}{\alpha_+}\Omega^{2n}
\right|+
\sum_{n=1,3,5,\ldots}^{\infty}
\ln\left|1+e^{2i\theta_{A,j}}\Omega^{2+4n}-e^{i\theta_{A,j}}\frac{Z}{\alpha_+}\Omega^{2n}\right|
\right)
\nonumber\\
&+&\frac{1}{\pi}
\frac{\sigma_H}{\sigma_+\sigma_-+\sigma^2_H}\left(
\sum_{n=0,2,4,\ldots}^{\infty}
\sum_{j=1}^{n_D}I^{D}_j\arg\left(1+e^{-2i\theta_{B,j}}\Omega^{2+4n}-e^{-i\theta_{B,j}}\frac{Z}{\alpha_+}\Omega^{2n}\right)-
\sum_{j=1}^{n_S}I^{S}_j\arg\left(1+e^{-2i\theta_{A,j}}\Omega^{2+4n}-e^{-i\theta_{A,j}}\frac{Z}{\alpha_+}\Omega^{2n}\right)
\right.\nonumber\\
&+&\left.\sum_{n=1,3,5,\ldots}^{\infty}\sum_{j=1}^{n_D}I^{D}_j
\arg\left(1+e^{2i\theta_{B,j}}\Omega^{2+4n}-e^{i\theta_{B,j}}\frac{Z}{\alpha_+}\Omega^{2n}\right)-
\sum_{j=1}^{n_S}I^{S}_j\arg\left(1+e^{2i\theta_{A,j}}\Omega^{2+4n}-e^{i\theta_{A,j}}\frac{Z}{\alpha_+}\Omega^{2n}\right)\right).
\end{eqnarray}
\end{widetext}
The expression for extended source/drain can be found by treating $I^{S/D}_j$ as infinitesimal and then converting the Riemann sum into an integral.
The obtained expression can now be used to fit measurements with multiple voltage probes for arbitrary current source and drain placed on the perimeter of the disk.

\acknowledgments
I wish to express sincere gratitude to Prof. J.I.A. Li for sharing their unpublished results and for his encouragement to publish this article. I would also like to thank Prof. Jian Kang for going over the calculations in the manuscript.
O.~V.~is supported by NSF DMR-1916958 and is partially funded by the Gordon and Betty Moore Foundation's EPiQS Initiative Grant GBMF11070, National High Magnetic Field Laboratory through NSF Grant No.~DMR-1157490 and the State of Florida.

\end{document}